\documentstyle[aps,prl,preprint,floats,epsfig]{revtex}  

\begin{document}

\preprint{\tighten\vbox{\hbox{\hfil CLNS 98/1544}
                        \hbox{\hfil CLEO 98-4}
}}


\title{
 \quad\\[1cm] \Large
Observation of High Momentum $\eta^\prime$ Production in $B$ Decay}

\author{CLEO Collaboration}

\maketitle

\tighten

\begin{abstract}

We report the first
observation of  $B\to \eta^\prime X$ transitions with high momentum
$\eta^\prime$ mesons. We observe
$39.0\pm 11.6$ $B$ decay events with $2.0< p_{\eta^\prime} < 2.7$ GeV/c,
the high momentum region where background from $b\to c$ processes
is suppressed.  We discuss the physical interpretation
of the signal,  including the possibility 
that it is due to $b\to s g^*$ transitions. Given that interpretation,
we find 
${\cal B}(B\to \eta^\prime X_s) = 
(6.2 \pm 1.6(stat) \pm 1.3(sys)^{+0.0}_{-1.5}(bkg))\times 
10^{-4} $ for $2.0<p_{\eta^{\prime}}<2.7$ GeV/c. 

\end{abstract}

{\renewcommand{\thefootnote}{\fnsymbol{footnote}}

\begin{center}
T.~E.~Browder,$^{1}$ Y.~Li,$^{1}$ J.~L.~Rodriguez,$^{1}$
T.~Bergfeld,$^{2}$ B.~I.~Eisenstein,$^{2}$ J.~Ernst,$^{2}$
G.~E.~Gladding,$^{2}$ G.~D.~Gollin,$^{2}$ R.~M.~Hans,$^{2}$
E.~Johnson,$^{2}$
% ...
\end{center}
 
\small
\begin{center}
$^{1}${University of Hawaii at Manoa, Honolulu, Hawaii 96822}\\
$^{2}${University of Illinois, Urbana-Champaign, Illinois 61801}\\
% ...
$^{13}${Stanford Linear Accelerator Center, Stanford University, Stanford,
California 94309}\\
% ...
$^{20}${University of California, San Diego, La Jolla, California 92093}\\
% ...
\end{center}

\setcounter{footnote}{0}
\newpage

%
%
%


Decays of the type $b\to s g^*$, gluonic penguins, are likely to
be important in future studies of $CP$ violation.
Gluonic penguin modes will be used to search
for direct $CP$ violation, and could complicate the interpretation of
some measurements of indirect $CP$ violation.
CLEO has reported the observation of signals in 
$\bar{B}^0\to K^-\pi^+$\cite{kpiprl}
and $B^-\to K^-\eta^\prime$\cite{etakprl}, exclusive modes 
which are expected
to be dominated by the gluonic penguin amplitude.
The inclusive decay $B\to \eta^{\prime} X$, where 
the collection of particles $X$
contains a single $s$ quark, 
is another signature of $b\to s g^*$ (followed by $g^*\to u \bar{u}$,
$d \bar{d}$ or $s \bar{s}$).  
Here we report the observation of 
the inclusive process $B\to \eta^\prime X$ 
and examine these data for evidence of $b\to s g^*$.

The data sample used in this analysis was
collected with the CLEO II detector at the
Cornell Electron Storage Ring.
This detector is designed to measure 
charged particles and photons with high efficiency and precision\cite{NIM}.
The data sample has an integrated luminosity of 
$3.1$ fb$^{-1}$ and contains $3.3 \times 10^{6}$ 
$B \bar{B}$ pairs.
Another data sample with an integrated luminosity of $1.6 $ fb$^{-1}$
was taken at an energy 60 MeV below the $\Upsilon(4S)$ resonance 
and is used to subtract the continuum background.

To isolate the signal and
reduce the large background from continuum production of $\eta^\prime$ mesons, 
we apply the $B$ reconstruction technique
that was previously used to isolate an 
inclusive signal for $b\to s\gamma$\cite{bsgamma}. This technique selects
$B\to \eta^{'} X$ events
in which the decay products of $X$ include a charged
kaon candidate in order to enhance the probability
of observing $b\to s g^*$.

We search for
$\eta '$'s with laboratory momenta in the range $2.0 < p <2.7$ GeV/c,
beyond the kinematic limit for most $b\to c$
decays. This range corresponds to a
region in $X$ mass from zero to 2.5 GeV.
In this momentum range we should be sensitive to $b \to s g^*$.
However, $b \to u$ decays with
$\eta^\prime$ mesons, such as $B^- \to \pi^- \eta^\prime$, and color-suppressed
$b \to c$ decays, such as $\bar B^0 \to D^0 \eta^\prime$, also populate this
interval.  Methods for discriminating among these possibilities will
be discussed later.

Events are selected using standard criteria for hadronic final states, and
we consider well-reconstructed tracks and photons.
Candidate $\eta^\prime$ mesons are then reconstructed in the 
$\eta^\prime\to \eta\pi^+ \pi^-$, $\eta\to \gamma\gamma$ mode.
For each $\eta$ candidate, the
$\gamma\gamma$ invariant mass 
must be within $30$ MeV of the nominal
$\eta$ mass. The $\eta$ candidate is fit kinematically to the
$\eta$ mass and is then combined with the charged pions
to form the $\eta^\prime$ candidate.

\begin{figure}
\centering
\mbox{\psfig{figure=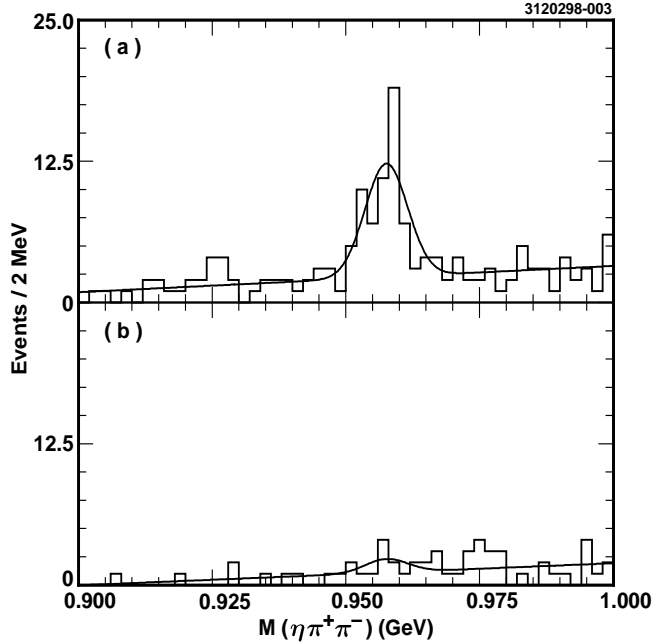,width=3.4in,height=3.4in}}
\label{brecon_fig1}
\centering
\caption{The distribution of $\eta\pi^+\pi^-$ mass for (a) on-resonance
data and (b) below-resonance data}
\end{figure}

We then form 
combinations of a charged kaon, an $\eta^\prime$ and $n$ pions
where $n\le 4$ (at most one of these pions is allowed
to be neutral).   
Eight decay modes and their charge conjugates are considered:
$B^-\to K^-\eta^\prime$,
$\bar{B}^0\to K^- \eta^\prime \pi^+$, $B^-\to K^-\eta^\prime\pi^0$,
$B^- \to K^- \eta^\prime \pi^-\pi^+$,  
$\bar{B}^0\to K^-\eta^\prime\pi^+\pi^0$,
$\bar{B}^0\to K^-\eta^\prime \pi^+\pi^-\pi^+$,
$B^-\to K^-\eta^\prime \pi^+\pi^-\pi^0$ and
$\bar{B}^0\to K^-\eta^\prime \pi^+\pi^-\pi^+\pi^0$.
For the charged kaon candidate 
we require that $dE/dx$ be within 
three standard deviations of the expected value.
We then require that these 
combinations be consistent in beam-constrained
mass ($M_B$) and energy difference
($\Delta E= E_{observed}- E_{beam}$) with a $B$ meson. 
(Here $M_B$ denotes the invariant mass with the energy
constrained to the beam energy.) 
We require $|\Delta E|<0.1$ GeV and $M_B>5.275$ GeV. 
In case of ambiguous hypotheses
we choose the best candidate in each event
on the basis of a $\chi^2$ formed from 
$M_B$ and $\Delta E$.

Following reference\cite{bsgamma}, we suppress the jet-like
continuum relative to the spherical $B\bar{B}$ events with 
requirements on $R_2$, (the ratio of second to zeroth Fox-Wolfram
moments) and $\theta_{S}$ (the angle between the sphericity
axis of the $B$ candidate and the sphericity axis of the remainder
of the event). $R_2$ is large for jet-like events and small for
spherical events. The variable $\cos\theta_S$ is isotropic for
signal events and peaks near $\cos\theta_{S}=\pm 1$ for continuum.
We require $R_2<0.45$ and $|\cos\theta_S|<0.6$.

The $\eta\pi^+\pi^-$ mass spectrum in the
high momentum window $2.0 <p_{\eta^\prime} <2.7$ GeV/c 
for the on-resonance
and off-resonance samples is shown in Fig.~1.
A fit to the $\eta^{\prime}$ peak
finds $50.7\pm 8.6$ events on the $\Upsilon(4S)$ resonance
 and $6.1\pm 4.1$ off resonance (unscaled). After accounting for the
differences in luminosity of the two samples,  
this gives an excess of $39.0\pm 11.6$ events.
Other ways of determining the yield give
consistent results.
We estimate a systematic error of $3\%$ from 
the uncertainty in the fitting procedure.

\begin{figure}
\centering
\mbox{\psfig{figure=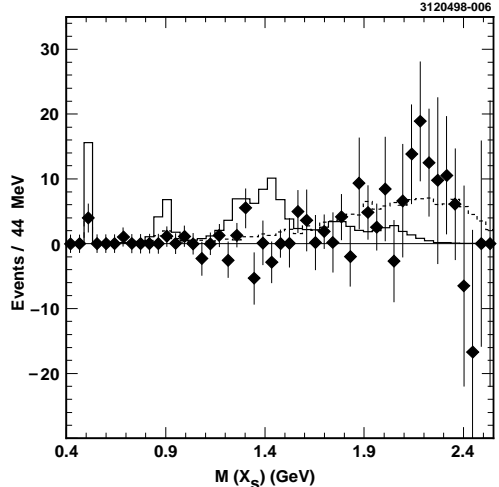,width=2.6in,height=2.6in}}
\centering
\caption{The continuum-subtracted
$M(X_s)$ distribution (points with error bars) with expected $M(X_s)$ 
distributions 
for a mixture of two-body $b\to s q \bar{q}$ (solid) and 
three-body $b\to s g^*$ with $g^* \to g\eta^\prime$ simulated
using JETSET
(dashed). The data points have been corrected
for the $M(X_s)$ dependent efficiency. Each simulation has been
normalized to the data yield.}
\label{xs_vs_d}
\end{figure}

We now study the invariant mass spectrum $M(X_s)$ of the particles recoiling 
against the $\eta^\prime$.
The continuum-subtracted $M(X_s)$ distribution
is shown in Fig.~2 and tabulated in Table~\ref{xsyield}.
The peak at the kaon mass
due to the exclusive decay mode $B^-\to \eta^\prime K^-$\cite{etakprl},
accounts for about 10\% of the inclusive yield. 
There is no significant excess in the $K^*$ mass region.
The remainder of the yield
comes from events with $X_s$ mass near or above charm threshold.
 Five sources can contribute to this distribution:
$\eta^\prime$ from secondary decays $b \to c$, $c \to \eta^\prime$;
color-allowed $b \to c$; color-suppressed $b \to c$; $b \to u$; 
and $b \to s g^*$.

\begin{table}[htb]
\caption{Yield in $M(X_s)$ bins for $B\to \eta^\prime X_s$.
The off-resonance yields must be scaled by 1.908 to account for
differences in energy and luminosity.
The detection efficiency drops as $M(X_s)$ approaches
 2.5 GeV because of the $\eta^\prime$ momentum cut.}
\medskip
\label{xsyield}
\begin{tabular}{llll}
$M(X_s)~(GeV)$  & N (on) & N(off) & Yield  \\ \hline 
$0.4<M(X_s)<0.6$ &  4  & 0 & $4\pm 2$ \\
$0.6<M(X_s)<1.2$ & $2.7\pm 2.1$  & $0.6\pm 1.1$ & $1.6\pm 2.9$  \\
$1.2<M(X_s)<1.8$ & $18.0\pm 4.9$  & $6.6\pm 3.2$ & $5.4\pm 7.6$  \\
$1.8<M(X_s)<2.5$ & $26.0\pm 6.4$  & $-0.8\pm 2.3$ & $27.5\pm 7.8$  
\end{tabular}
\end{table}

    Secondary decays have been reliably 
simulated with the Monte Carlo program.
These include processes such as
$\bar B^0 \to D^+ \pi^-$, $D^+ \to \eta^\prime \pi^+$ and
$\bar B^0 \to D_s^- D^+$, $D_s^- \to \pi^- \eta^\prime$.  The yield
from secondary sources is thus estimated to be less than 0.2 events.

We have also considered the possibility of color-allowed $b\to c$
backgrounds such as $B\to D\eta^\prime\pi$.
We have searched for this decay in a lower $\eta^\prime$ momentum
range, modeling the decay with 3-body phase space.
This search gives an upper limit of ${\cal B}(B\to
D\eta^{\prime}\pi)<1.3\times 10^{-3}$, 
corresponding to a background of fewer than 1.4 events
in the signal region. Thus, neither secondary decays nor
color-allowed $b \to c$ decays are a 
significant source of the high momentum $\eta^\prime$ signal.


We next consider $b\to u$ modes.  First, we check for the
presence of an $s$ quark in the final state by forming
a $\chi^2$ difference based on $\Delta E$ and the
resolution-normalized $dE/dx$ residual for the candidate kaon.
The $\Delta E$ distribution for $b\to u$
modes is shifted above that for $b\to s$ modes
because the kaon mass is attributed to a pion.
A fit to this $\chi^2$ difference, using 
$B\to\eta^\prime\pi, \eta^\prime\rho, \eta^\prime a_1$
for the $b\to u$ contribution and a model of $b\to s g^*$
for the $b\to s$ contribution, gives the yields
$f(b\to s g^*) = 82\pm 20\% $ and $f(b\to u) = 18\pm 20\% $.

Further information on $b\to u$ comes from the $M(X_s)$
distribution.
The dominant $b\to u$ modes with an $\eta'$ are expected 
to be $B\to\eta^\prime\pi, \eta^\prime\rho$ and $\eta^\prime a_1$.
These modes have $X_s$ mass below 1.8 GeV, where we see no strong evidence
for a signal.  
The theoretical expectation for $b\to u$,\cite{atwood} or 
summing the exclusive modes which may contribute\cite{cheng},
is $(3.5-7.0)\%$ of the signal yield.  The contribution
with $M(X_s)>1.8$ GeV and $2.0 < p_{\eta^\prime}<2.7$ GeV
is likely to be much smaller.

Experimental searches for the color-suppressed
$b \to c$ modes $\bar B^0 \to D^{(*)} \eta^\prime$
\cite{color}\cite{dstpi}, while showing no evidence for them, place 
unrestrictive upper limits.  Theoretical expectations for the 
branching fractions for these modes are in the range
$(1.5 - 6.0) \times 10^{-5}$ \cite{neubert}, implying a yield of high momentum
$\eta^\prime$ of (2.1 -- 8.6)\% of the observed yield.

To search for these modes in the data, we examine the $M(X_s)$
distribution for neutral modes.  The mode
$\bar B^0 \to D^0 \eta^\prime$ has a spike at the
$D^0$ mass, while that for 
$\bar B^0 \to D^{*0} \eta^\prime$ has a broader peak
at the $D^{*0}$ mass as shown in Fig.~3.
This distribution limits the contribution of
$\bar B^0 \to D^{0} \eta^\prime$ to 15\% of the signal. 

\begin{figure}
\centering
\mbox{\psfig{figure=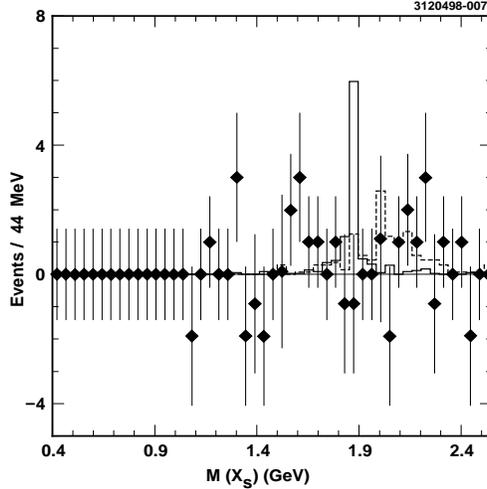,width=2.6in,height=2.6in}}
\centering
\caption{The continuum-subtracted 
distribution of $X_s$ mass for neutral modes with the
expectations from $\bar{B}^0\to D^0\eta^{'}$ (solid) and
$\bar{B}^0\to D^{*0}\eta^\prime$ (dashed) superimposed.
The simulations have been normalized to the data yield.}
\label{neutral_Xs}
\end{figure}

Information about $\bar B^0 \to D^{*0} \eta^\prime$
comes from the mass distribution obtained by removing
a single pion (charged or neutral) from the $X_s$ system, such that the
remaining particles are consistent with coming from a $D^0$ decay.
$\bar B^0 \to D^{*0} \eta^\prime$ peaks sharply at the $D$ mass 
in this distribution 
as shown in Fig.~\ref{pseudod}.  The absence of such a peak in the data
limits $\bar B^0 \to D^{*0} \eta^\prime$ to 26\% of the signal.
The limits on $D^0\eta^{'}$
and $D^{*0}\eta^{'}$
constrain the total contribution of color suppressed decay 
to be less than 41\% of the signal.

We have also tried
to describe the data in Fig.~2 with a combination of
$\bar B^0 \to D^0 \eta^\prime$, $\bar B^0 \to D^{*0} \eta^\prime$, and
$\bar B^0 \to D^{**}(2.2) \eta^\prime$ 
($D^{**}(2.2)$ being a hypothetical broad state decaying 
into $D \pi$ and $D^* \pi$), and have found no combination with
a confidence level above 2.7\%.  We conclude that while these modes
could contribute to our signal, they are unlikely to account
for it fully.

\begin{table}[htb]
\caption{Detection efficiency for $B\to \eta^\prime X_s$ modes}
\medskip
\begin{tabular}{ll}
Mode & $\epsilon$  \\ \hline 
$B\to K\eta^\prime$ &  $ 0.076\pm 0.006$ \\
$B\to K^{*}(892)\eta^\prime$ & $0.058\pm 0.005$ \\
$B\to K_1(1270)\eta^\prime$  & $0.050\pm 0.005$\\
$B\to K_1^{*}(1406)\eta^\prime$ & $0.053\pm 0.005$\\
$B\to K_2^{*}(1429)\eta^\prime$ & $0.051\pm 0.005$\\
$B\to K_3^{*}(1774)\eta^\prime$ & $0.046\pm 0.005$ \\
$B\to K_4^{*}(2200)\eta^\prime$ &  $ 0.046 \pm 0.005$ \\ 
$B\to D^0\eta^\prime$ &  $ 0.025 \pm 0.002$ \\ 
$B\to D^{*0}\eta^\prime$ &  $ 0.026 \pm 0.002$ \\ 
$B\to D(2.2)\eta^\prime$ &  $ 0.011 \pm 0.003$ \\ \hline
Equal mixture of exclusive $b\to s g^*$ modes  &   $0.055\pm 0.003$ \\
$B\to \eta^\prime s \bar{d}, \eta^\prime s \bar{u}$
(JETSET hadronization) & $0.055\pm 0.003$
\label{efficiency}
\end{tabular}
\end{table}

Finally, we consider $b \to s g^*$.
Conventional $b \to s q \bar q$ operators predict
an $X_s$ mass distribution that peaks near 1.5 GeV.
This description fits the $M(X_s)$ spectrum poorly (1\% C.L.).  However, the
process $b \to s g^*$ with $g^* \to g \eta^\prime$ from the QCD anomaly, 
which has
the attractive feature that it accounts for the large mass of the
$\eta^\prime$ relative to other members of its SU(3) multiplet,
gives a three-body $g s \bar q$ mass spectrum 
that peaks above 2 GeV\cite{atwood},\cite{hou}--\cite{du}
as shown in Fig.~2.
A fit of this model to the data gives a C.L.
in the 25 - 72\% range.

In what follows, we
shall compute the $B\to X_s \eta^\prime$ branching fraction assuming
that it is due to $b\to s g^*$, and allow for a background from
$D^{(*(*))}\eta^\prime$ assuming that these decays occur at
rates consistent with standard expectations.  It is also possible,
however, that $D^{(*(*))}\eta^\prime$ is occuring at a rate
that is up to five times larger than expectation; if so, 
half our signal could be from these modes.

\begin{figure}
\centering
\mbox{\psfig{figure=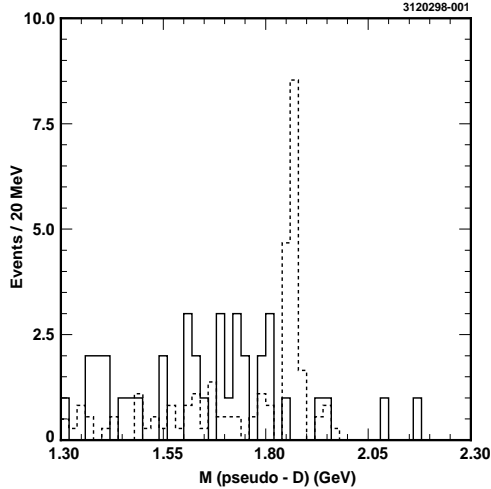,width=2.6in,height=2.6in}}
\centering
\caption{The continuum-subtracted 
distribution of pseudo-$D$ mass (solid) with the
expectation from $\bar{B}^0\to D^{*0}\eta^\prime$ (dashed).
The simulation has been normalized to the data yield.}
\label{pseudod}
\end{figure}

The detection
efficiency for $b\to s g^*$ and for the color-suppressed $b\to c$
decays are listed in Table II. 
We observe that the efficiency for the $b\to c$ decay mechanisms
is half that for
$b\to s g^*$, so the computed $B\to X_s \eta^\prime$ branching fraction
depends on our assumption that the signal is $b \to s g^*$.
The $b\to s g^*$ decays are modeled by allowing the JETSET 
Monte Carlo to hadronize the $s$ quark and gluon. We also 
generate a number of exclusive $b\to s g^*$ decay modes.
The average efficiency
for exclusive modes with equal weights is equal to the JETSET efficiency. 
The  detection efficiency is averaged over 
$B^0$ and $B^+$ mesons and corrects
for unobserved modes with $K^0$ mesons but does not include 
$\eta^{\prime}$ branching fractions.  
To determine the  uncertainty in efficiency due to 
the modeling of the signal, we vary 
the relative weights of different modes (increasing the fractions
of $K_3^*\eta^{\prime}$ and $K_4^*\eta^{\prime}$ to 50\%); this 
leads to a systematic error of 16\%. 
No attempt has been made to calculate the
branching fraction for decays that lie outside the $\eta^\prime$
momentum window, as such a calculation would
be extremely model dependent.

The dominant source of experimental systematic error is 
due to the modeling of the $X_s$ system\cite{highXs}. 
Other sources include the choice of background parameterization
and the uncertainty in the tracking and photon detection.
We have also included a second systematic error for the 
possible contribution of 
color suppressed $b\to c$ modes. This is determined by using the
largest model prediction for these modes and taking into account 
their lower acceptance. The theoretical predictions
are multiplied by 1.5 as an estimate of the 
theoretical uncertainty. 
Assuming an average detection efficiency of 5.5\%, 
appropriate for $b\to s g^*$, we obtain
${\cal B}(B\to \eta^\prime X_s) = 
(6.2 \pm 1.6(stat) \pm 1.3(sys)^{+0.0}_{-1.5}(bkg))\times 
10^{-4} $ for $2.0<p_{\eta^\prime} <2.7$ GeV/c. 

A number of interpretations have been proposed to account
for the large branching fraction of $B\to \eta^{\prime} X_s$.
These include: (I) conventional $b\to s q\bar{q}$ operators with 
constructive interference between 
the $u\bar{u}, d \bar{d},$ and $s \bar{s}$
components of the $\eta^\prime$ \cite{datta,lipkin}, 
(II) $b\to c\bar{c} s$
decays enhanced by $c \bar{c}$ content in the $\eta^\prime$ wavefunction
\cite{halperin,chao}, 
and 
(III) $b\to s g^*, g^*\to g \eta^\prime$ from the $\eta^\prime$ QCD 
anomaly \cite{atwood},
\cite{hou}--\cite{du}. The observed branching fraction is larger
than what is expected from Scenario I. Furthermore,
scenarios I and II will give an $X_s$ mass distribution 
peaked near 1.5 G$e$V.  Only 
scenario III gives a three-body $ g s \bar{q}$ $X_s$ mass spectrum 
that peaks above 2 GeV.

We have also searched
for high momentum $B\to \eta X_s$ decays, with $\eta\to \gamma\gamma$
and $\eta\to \pi^- \pi^+ \pi^0$ and $2.1<p_{\eta}<2.7$ GeV.
In the $\eta\to\gamma\gamma$ mode, we observe
$164\pm 17$ events on resonance, $50\pm 10$ events below resonance.
The expected $b\to c$ background is $14\pm 8$ events. This gives
a net yield of $54\pm 26$ events.
In the $\eta\to \pi^-\pi^+ \pi^0$ mode, we obtain
an on-resonance yield of $52.4\pm 10.5$  and a below-resonance
yield of $24.9\pm 6.7$ events. This corresponds to an
excess of $4.9\pm 16.5$ events.
The limit obtained by combining the two $\eta$ decay modes 
and allowing for systematic uncertainty is 
$ {\cal B}(B\to \eta X_s)<4.4\times 10^{-4}$.
The theoretical  expectation is 
that this rate will be suppressed relative to $B\to \eta^\prime X_s$ by 
$\tan^2\theta_P$\cite{atwood,hou}
where $\theta_P$ is the pseudoscalar mixing angle, which is
consistent with our result.

In summary, we have observed a signal of 39.0 $\pm$ 11.6 events in high
momentum $B \to \eta^\prime X_s$ production.  
The interpretation $b \to s g^*$ is
consistent with all features in the data.
Given that interpretation, the branching fraction is
${\cal B}(B\to \eta^\prime X_s) = 
(6.2 \pm 1.6(stat) \pm 1.3(sys)^{+0.0}_{-1.5}(bkg))\times 
10^{-4} $ for $2.0< p_{\eta^\prime} <2.7$ GeV/c.

We gratefully acknowledge the effort of the CESR staff in providing us with
excellent luminosity and running conditions.
This work was supported by 
the National Science Foundation,
the U.S. Department of Energy,
Research Corporation,
the Natural Sciences and Engineering Research Council of Canada, 
the A.P. Sloan Foundation, 
and the Swiss National Science Foundation.


\begin{thebibliography}{99}
\bibitem{highXs}{As a consistency check,
we have also evaluated the signal yield without the
$\eta^\prime$ momentum cut in order to include a larger $M(X_s)$ range.
We have attempted to evaluate the 
yield for $2.5<M(X_s)<3.0$ GeV and find $26.9\pm 11.0(stat)$
signal events after background subtraction. 
Due to the large systematic uncertainty in the $b\to c$ subtraction,
we do not determine a branching fraction for this region.
In the range $0.4<M(X_s)<2.5$ GeV, we find 
a branching fraction ${\cal B}(B\to \eta^\prime X_s)=
(5.9\pm 1.9\pm 1.3)\times 10^{-4}$ for $0.4<M(X_s)<2.5$ GeV
in good agreement with the primary result given in this paper.}
\end{thebibliography}
\end{document}